\documentclass[sigconf]{acmart}

\usepackage{booktabs} 
\usepackage{color}
\usepackage{comment}

\setcopyright{rightsretained}
\usepackage{algorithm}
\usepackage{algpseudocode}
\acmDOI{10.475/123_4}

\acmISBN{123-4567-24-567/08/06}

\acmConference[]{Report}{}{}
\acmYear{2017}
\copyrightyear{2017}

\acmPrice{15.00}

\begin{document}
\title{Dynamic Intention-Aware Recommendation System}

\author{Shuai Zhang}
\orcid{1234-5678-9012}
\affiliation{%
  \institution{University of New South Wales}
  \streetaddress{K17, CSE, UNSW}
  \city{Sydney}
  \state{NSW}
  \postcode{2052}
    \country{Australia}
}
\email{shuai.zhang@student.unsw.edu.au}

\author{Lina Yao}
\affiliation{%
  \institution{University of New South Wales}
  \streetaddress{K17, CSE, UNSW}
  \city{Sydney}
  \state{NSW}
  \postcode{2052}
  \country{Australia}
}
\email{lina.yao@unsw.edu.au}

\begin{abstract}
Recommender systems have been actively and extensively studied over past decades. In the meanwhile, the boom of Big Data is driving fundamental changes in the development of recommender systems. In this paper, we propose a dynamic intention-aware recommender system to better facilitate users to find desirable products and services. Compare to prior work, our proposal possesses the following advantages: (1) it takes user intentions and demands into account through intention mining techniques. By unearthing user intentions from the historical user-item interactions, and various user digital traces harvested from social media and Internet of Things, it is capable of delivering more satisfactory recommendations by leveraging rich online and offline user data; (2) it embraces the benefits of embedding heterogeneous source information and shared representations of multiple domains to provide accurate and effective recommendations comprehensively; (3) it recommends products or services proactively and timely by capturing the dynamic influences, which can significantly reduce user involvements and efforts.
\end{abstract}

%
%


\maketitle

\section{Introduction}

Recommender System is a well-recognized and efficient way to provide end users with personalized services and products, overcome information overload and boost sales. Although various recommendation techniques such as collaborative filtering, content-based filtering and  context-aware recommendation system, have achieved tremendous success in real-world applications in the past decades, most existing approaches still remain insufficient for the following situations: (1) CF-based recommender systems work on historical data to learn user preferences, they are unable to deliver a satisfying recommendations while user preferences change over time or only can grab partial information to make inaccurate inference. (2) hybrid models such as integrating content-based features can deal with cold-start and sparse user-item data to some extent, but they still have limitations in handling the situation wherein user's desires don't match inferred preferences.

In this paper, we introduce a novel concept to tackle aforementioned challenges, and denote it \textit{dynamic intention-aware recommendation system}. To the best of our knowledge, such type of recommender system has not been explored much in literature. For example, in the book recommendation case, the system will produce a list of books for users according to their purchasing history or browsing activity. But, in reality, the user may be looking for some new released books in the field that she will major in. Without understanding the user needs correctly, the recommendation system is not capable of recommending properly.

To recognize user intention, we may need information from other fields/domains. In the aforementioned scenario, we may find that the user will major in computer science in a university from her social media posts. Thus, it is beneficial to leverage these available social network data to assist in recognising user intentions. In addition, we may predict that the user will possibly intend to watch the movie \textit{Doctor Strange} after observing that the user gave a high rating for the comic book \textit{Doctor Strange}, or shared relevant online information etc. In such case, the proposed intention-aware recommender system is able to leverage the multi-domain information and relationship to proactively recommend her the movie \textit{Doctor Strange} while her GPS records show that she is hanging around a cinema.

Although existing recommender systems have considered user ratings, user comments on items, social influence, user cross-domain preference and personality \cite{li2009transfer, ekstrand2016behaviorism, loni2014cross, elkahky2015multi}, they either needs user participation, e.g., initiating a request, or is incapable of recommending in a timely fashion. This reactive fashion sometime causes frustration or interruptions~\cite{sabic2016proactive} and is unable to capture user real needs. By anticipating user intentions and demands beforehand, we can reduce user involvements appropriately and minimize the user efforts. Our proposed framework enables the recommender system to proactively direct users to the products and services she desired. For instance, recommendation system is supposed to push a notification actively about nearby restaurants user may like at mealtimes, rather than wait for user to send an explicit or implicit request, e.g, browsing certain products. It is noted that the proposed intention-aware recommender system will not bug user as it only make proactive move by telling user true desires.

Unlike content-aware recommendation, which incorporates the contextual information to provide items to users under certain circumstances~\cite{adomavicius2015context}. Our proposed dynamic intention-aware recommendation system can be considered as a step further, it takes the contextual information into account, furthermore, it will unearth user intentions from historical interactions and footprints from social media and physical world (e.g., Internet of Things) before producing recommendations. It
aims to provide recommendations in a comprehensive manner by fully utilizing abundant auxiliary information from multiple domains.

The rest of this paper is organized as follows. The research questions are given in Section 2.  A brief review of related work is shown in Section 3. Then, the proposed methodology and system structure of the dynamic intention-aware recommendation system are presented in Section 4. Section 5 illustrates the proposed experiments. Conclusion is provided in Section 6.


\section{research questions}
The general research question to be addressed in our proposed work is \textit{How to design and develop a dynamic intention-aware recommendation model across multiple domains}. To solve this question, we decompose it into several sub-questions. Details are as follows:

\subsubsection{How to identify user intentions.} Intention prediction plays a role in enhancing recommendation performance. However, it is nontrivial task to accurately understand user intentions. One of the main challenges would be that user's real intention usually is tangled with a set of complex user behaviors and interactions in reality. Multi-dimensional and multi-scale online and offline data would be greatly helpful to understand the complexity to discern the underlying intention. Recommender systems equipped with intention prediction could be able to satisfy users needs more considerately and accurately.

\subsubsection{How to implement the cross-domain recommendation system with user intentions integrated.} To the best of our knowledge, very few works aim to solve this problem. Additionally, in the intention-aware recommendation system, recommending task is processed among multiple domains. Even though the cross-domain recommendation system has been actively studied, it is still an underexplored topic and remains a challenge ~\cite{Ricci:2010:RSH:1941884}. Thus, we need to devise a comprehensive approach to effectively leverage user intentions with cross-domain recommendations.

\subsubsection{How to embed the heterogeneous source information.} Abundant auxiliary information is usually available in cross-domain recommendation system. These rich information can be leveraged to boost the performance. The heterogeneous information, such as, structural content, textual content and visual content, cannot be dealt with uniformly. Different methods should be applied to embed these diverse information into the recommender model.

\subsubsection{How to deal with dynamic influence.} Dynamic influence includes sequential influence and temporal influence. Generally, user intentions exhibit sequential patterns and variations from time to time. For instance, the check-in information tends to have a sequential property along with periodical pattern, we can anticipate a user's next destination from her historical traces. The temporal dynamics, which have been explored frequently~\cite{koren2010collaborative,lathia2010temporal},  also pose great influence on user intentions. User intentions and preference may be drifting with weekly changes, monthly changes, and seasonal changes. Modeling such changes along temporal dimension also is a challenging task due to the complexity in capturing temporal patterns. In our work, we propose a dynamic framework to enable an intention-aware recommender system over time.

\section{Related Work}
In this section, we describe several research fields that are similar with or related to our researches.

\begin{figure*}
\label{comptraintime}
\includegraphics[width=1\textwidth]{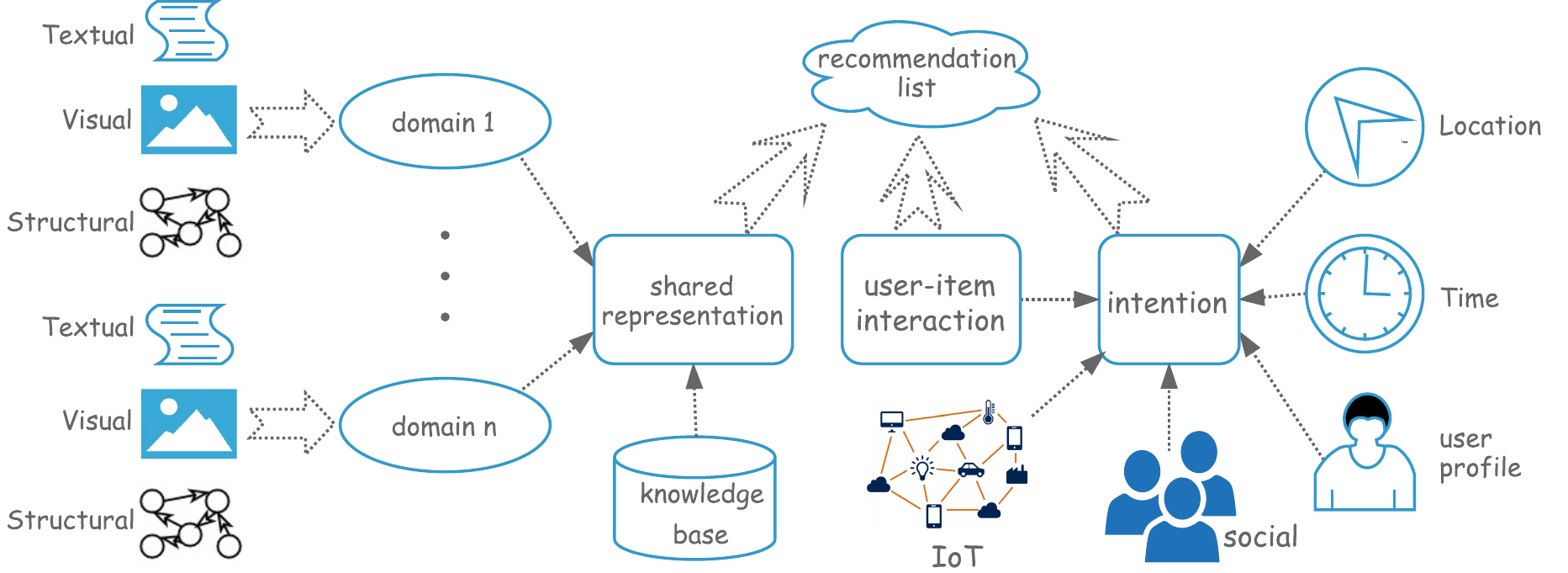}
\caption{System Architecture of Dynamic Intention-aware Recommendation System}
\vspace{0.5cm}
\end{figure*}

\subsection{Intention Mining}

Intention mining~\cite{khodabandelou2013supervised,khodabandelou2013process} is a technique aimed to uncover user intentions from their historical interactions or behaviours. It has been widely studied in many research fields, such as, web search, social media, multimedia and robotics.

There are different approaches for mining user intentions. Mei et al.~\cite{mei2007modeling} presented a method for classifying the \textit{capture intention} of home videos into seven psychological categories. Sun et al.~\cite{sun2015mining} presented a method for anticipating user location intention from their mobile search queries. Ding et al.~\cite{ding2015mining} proposed a domain adaptive convolutional neural network to identify user implicit consumption intention from social media. Castellanos
et al.~\cite{castellanos2012intention} devised an intention analysis tool named \textit{intention insider} to extract user intentions in the online forums. \cite{khodabandelou2013supervised,aarno2008motion} utilized the hidden markov model (HMM) for recognizing the intentions underlying user activities. He et al.~\cite{he2016intention} proposed an Recurrent Neural Network (RNN) based methods to capture user queries intention for video browsing task.  Yao et al.\cite{yao2015attention} use RNN to model the dynamics of intention process in conversation machines.


\subsection{Cross-Domain Recommendation System}
Most of the existed recommendation systems focus on single domain. However, large e-commerce companies often receive feedback from multiple domains. Meanwhile, it has triggered an increasing interest in cross-domain recommendation system. Li et al.~\cite{li2011cross} categorized the domain into three types: system domain, which refers to type of items (e.g. books, movies, tweets); Data domain, which is split by the types of feedback (e.g. implicit or explicit); Temporal domain, which is divided according to the rating timestamps. Here, we mainly explore the system domain, and consider the data domain and temporal domain as side information.

Techniques to solve the cross-domain recommendation system are quite different from which deal with the single-domain. In~\cite{li2009transfer}, the authors proposed a rating-matrix generative model for cross-domain collaborative filtering by clustering the related rating matrices. However, this method is computational expensive and is hard to learn a good cluster-level rating model. In~\cite{elkahky2015multi}, the authors proposed a multi-view deep learning model to generate recommendation from multiple domains with high scalability. Zhang et al.~\cite{zhang2016multi} applied active learning to cross-domain recommendation system introducing both domain-specific and domain-independent knowledge.


\subsection{Dynamics in Recommendation System}

Dynamics in recommendation system mainly consist of temporal status and spatial information. In the intention-aware recommendation system, we will mainly concentrate on these two facets of dynamics. Dynamics both existed in users and items. For example, user preferences change over time; and the popularity and click-through rates of new articles tend to decay with time.

Several works have been done to model the dynamics in recommendation system. Koren et al.~\cite{koren2010collaborative} presented a method to capture the time drifting of user preferences and applied it to both factorization model and neighborhood model. Chu et al.~\cite{chu2009personalized} proposed a feature-based bi-linear regression model, which can efficiently utilize the dynamic features to provided personalized recommendation and solve the cold-start problem. Yuan et al.~\cite{yuan2013time} developed collaborative recommendation model by incorporating temporal and geographical information. In~\cite{yao2016poi}, the authors leveraged the temporal matching between point-of-interest (POI) popularity and user regularity to improve the POI recommendation.

\section{methodology}
In this section, we present the details of how we overcome the problems mentioned in Section 2. Figure 1 illustrate the system architecture of the dynamic intention-aware recommendation system.

\subsection{Intention Prediction}
User intentions are determined by many factors. It is associated with user-item interactions, spatio-temporal dynamics, and user's social influence etc. We intend to develop a model that enables the recommendation system to capture user intentions from these multi-dimensional and heterogeneous raw data. We list several applicable techniques as follows.

\subsubsection{Intention Classification.} As mentioned before, most works treat the intention mining process as a classification problem. Before classifying intentions, they defined the types of intentions according to expertise knowledge. We can adopt the same method to identify intentions in the recommender system. To achieve the goal, several classifying methods, such as, regression tree, naive bayes classifier, etc. could be employed.
However, these approaches have two main shortcomings: firstly, they need prior knowledge such as intention types; secondly, they do not model the sequential property of intention. Therefore, methods that can capture the sequence features are more suitable. We have been investigating leverage non-parametric model with neural conditional random fields~\cite{do2010neural} to tackle the issues.

\subsubsection{Intention Inference from Real World}
Internet of Things (IoT) involves not only the connection and integration of devices that monitor the physical world parameters, which may be temperature, pressure, altitude, motion, proximity, biometrics, sound, images, etc.; But also it implies the aggregation, relationship, and analysis of the information those devices create in order to take action on the situation. The rising popularity of smart phones, built on advances of IoT, opens a new world of broader possibility and innovations with building more effective recommender system, for better understanding user's real desires and behavioral patterns.

Information scattered in various physical sources (ambient sensors, mobile phones etc.) in fact describes the inherent characteristics of the same user in various aspects, and hence their predicted results should be identical. The motivation behind this task lies in that different patterns may be unveiled in different domains, and these patterns may be complementary or mutually enhanced. Different source of data may show discrete partial views of a person's behaviors, in the meantime, like solving a puzzle, putting all data sources together will divulge the one. In particular, the person's behavior should show source disparity, as well as consistency. We propose to unite sparse dictionary learning and shared structure learning together for quantifying the shared and specific features in the subspaces to handle intra-class and inter-class behavior variability.

\subsubsection{Intention Inference from Social Data Analytics} Social networks data has been proved to be useful in improving the recommendation accuracy ~\cite{bao2012location}. In our work, we intent to leverage the tremendous amounts of content and linkage data as side information to anticipate user intentions. Typically, intentions on social media are classified into explicit and implicit intentions, while most intentions are implicitly expressed and user may not be entirely aware of~\cite{ding2015mining}. We will make use of both kinds of intentions, and focus on implicit intention mining in social network data.

\subsubsection{Dynamic Intention.} To model the dynamic intentions, we mainly investigate two techniques: HMM and RNN. HMM is an efficient dynamic tool for modelling sequential data. To apply HMM to Intention Mining, we consider the hidden states as the intentions behind the sequences of activities~\cite{khodabandelou2013supervised}, and then, discover the most likely intention given the observable traces of user activities. After having built the model, we can adopt a supervised or an unsupervised approach to estimate the parameters. Another deep learning model RNN can also deal with sequential data, the advantages of RNN over HMM is that RNN take into account the long-term dependencies~\cite{he2016intention}.

\subsection{Feature Extraction and Embedding}
To leverage the abundant multi-source content information, such as, attributes, text and photos, etc., from the user profiles as well as the item contents. Similar to \cite{zhang2016collaborative}, we divided all the content information into three types: textual information, structural information, and visual information.

\subsubsection{Textual Information} Textual information, such as, plot of a movie, abstract of a article and summary of a book, are very useful descriptive features in recommender systems\cite{wang2011collaborative,zhang2016collaborative}.  We propose to apply word embeddings atop auto-encoders, or its variants like denoising auto-encoders and marginalized auto-encoders, to learn useful feature representations from text. Auto-encoders is an unsupervised deep learning model that can learn high level representations and significantly reduce the feature dimensionality. For large collection of documents, topic modeling algorithms such as Latent Dirichlet allocation (LDA) can be used to discover the "topics" from the  text corpus. LDA will generate an interpretable low-dimensional representations from documents, compare with auto-encoders, it enables to retain more semantic properties in the representations.

\subsubsection{Structural Information} Structural information includes the attributes of items, and the relationships among items and users. In general, this information can be represented as graphs. Zhang et al. \cite{zhang2016collaborative} introduced a Bayesian form of TransR, a heterogeneous network embedding method, for interpreting structural informations. In \cite{yao2016things}, Yao et al. proposed a hyper-graph framework to model spatiotemporal correlations among users and things in Internet of Things (IoT) recommendation.

\subsubsection{Visual Information} Visual information affects user's decision to some degree. For example, an attractive book cover or movie poster will draw more customer's attentions.  To capture the visual features, such as, color, shape, size and texture, from images, the following two proposed methods can be adopted. Stacked convolution auto-encoders~\cite{zhang2016collaborative}, which combines stacked auto-encoders with convolution techniques, can be used to extract semantic representations from images; Two complementary techniques: LBP (local binary pattern) and HOG (histogram of oriented gradients), can also be jointly applied to capture the texture and shape information.

After feature representations, we will fuse them to our recommendation system. For some tasks, simply concatenation of heterogeneous features may achieve acceptable results.  While other tasks may require more effective combinatorial features, which can be archived by making use of a deep learning based model to automatically and optimally combine features.

\subsection{Recommendation}

\subsubsection{Shared Representations}In cross-domain recommender system, multiple domains are usually related with each other~\cite{li2009transfer}. Thus, the rich information from multiple domains could be pulled together to form shared representations~\cite{li2009transfer}, so as to alleviate the sparsity problem in cross-domain recommendation system.  To learn the shared representations, we mainly investigated two techniques: Multi-Task Learning and Multimodal Deep Learning.  Multi-Task learning is an approach that learns multiple tasks simultaneously by generating shared representations. It allows one task to make use of features from related task to improve accuracy and yield better generalization. The combination of Multi-Task learning with neural networks will further enhance performance~\cite{caruana1998multitask}. Multimodal deep learning is another shared representation learning approach. It is built over restricted boltzmann machine or auto-encoders, and enables to capture the correlations across different modalities.


\subsubsection{Knowledge Base} Knowledge base stores the domain expert knowledge. It can be utilized as a complementarity when item or user content information is too sparse. Various works~\cite{middleton2009ontology,trewin2000knowledge} have been done to leverage the expertise and inference capability to enhance recommendation system performance. In our work, we intend to employ some open-source knowledge base, such as, YAGO, DBpedia, and domain ontology for reasoning and further improving the multi-domain recommendation result.

Finally, the proposed recommendation system will provide users with an individual or a group of recommendations containing items from different domains to users proactively.

\section{Proposed Experiments}

\subsection{Dataset}
Another important consideration is the requirement of an appropriate dataset. Most existing datasets used for recommendation lie in a certain topics, such as MovieLens, LibraryThing etc. Some of them may integrate auxiliary information from other domains, e.g., MovieTweeting. We are planning to augment an existing large scale dataset by crawling from a broad range of sources, e.g., social medias, location based social networks and sensory data from city-scale ambient sensors. The dataset would be qualified for evaluating our approach by providing necessary micro and macro information fitting in our dynamic intention-aware recommender system. The dataset will be publicly available to contribute to the community.

\subsection{Evaluation}
Traditional recommendation systems mainly concentrate on improving accuracy. However, McNee et al.~\cite{mcnee2006being} argued that we shall recommend from the user-centric perspective to not only deliver recommendations accurately but also satisfactorily.

In this work, the proposed recommender system consists of two stages: intention mining and recommending. Thus, we need to evaluate systematically both the accuracy of the inferred intentions and the results of recommendation. To measure the intention mining results, same as~\cite{khodabandelou2013supervised}, we can choose \textit{recall} and \textit{precision} metrics, or the combination of them: \textit{F-factor}. For recommendation evaluation, we could adopt the commonly metrics (e.g.,RMSE, MAE, AUC, precision/recall) for different categories (e.g., prediction, ranking, classification) of recommendation tasks. In addition, we also consider a well-designed A/B test to evaluate the performance. A qualitative evaluation will also be conducted. We will take a questionnaires survey to gather the feedback from a wide range of end users for further analyzing and evaluating our approach.

\section{Conclusion}
In this paper, we present our proposal on a dynamic intention-aware recommendation system. We discussed the motivations and necessities of it. In order to build a dynamic intention-aware recommmder system, we divide this task into several subgoals and conduct extensive literature reviews on each part. With the increasing amounts of online and offline data available, it will be more achievable to discover user intentions accurately. The inferred intentions will also facilitate the recommendation system to provide better services as well as to improve user experiences.

\bibliographystyle{ACM-Reference-Format}
\bibliography{sample-sigconf.bbl}

\end{document}